\documentstyle[aps,preprint,pra,graphicx,url]{revtex}
\begin{document}
\newcommand{\etal}{{\em et al.}\/}
\newcommand{\IP}{inner polarization}
\newcommand{\IPF}{\IP\ function}
\newcommand{\IPFs}{\IP\ functions}
\newcommand{\auth}[2]{#1 #2, }
\newcommand{\oneauth}[2]{#1 #2,}
\newcommand{\andauth}[2]{and #1 #2, }
\newcommand{\book}[4]{{\it #1} (#2, #3, #4)}
\newcommand{\jcite}[4]{#1 #2(#4)#3}
\newcommand{\et}{ and }
\newcommand{\erratum}[3]{\jcite{erratum}{#1}{#2}{#3}}
\newcommand{\inbook}[5]{In {\it #1}; #2; #3: #4, #5}
\newcommand{\JCP}[3]{\jcite{J. Chem. Phys.}{#1}{#2}{#3}}
\newcommand{\jms}[3]{\jcite{J. Mol. Spectrosc.}{#1}{#2}{#3}}
\newcommand{\jmsp}[3]{\jcite{J. Mol. Spectrosc.}{#1}{#2}{#3}}
\newcommand{\jmstr}[3]{\jcite{J. Mol. Struct.}{#1}{#2}{#3}}
\newcommand{\cpl}[3]{\jcite{Chem. Phys. Lett.}{#1}{#2}{#3}}
\newcommand{\cp}[3]{\jcite{Chem. Phys.}{#1}{#2}{#3}}
\newcommand{\pr}[3]{\jcite{Phys. Rev.}{#1}{#2}{#3}}
\newcommand{\jpc}[3]{\jcite{J. Phys. Chem.}{#1}{#2}{#3}}
\newcommand{\jpcA}[3]{\jcite{J. Phys. Chem. A}{#1}{#2}{#3}}
\newcommand{\jpca}[3]{\jcite{J. Phys. Chem. A}{#1}{#2}{#3}}
\newcommand{\jpcB}[3]{\jcite{J. Phys. Chem. B}{#1}{#2}{#3}}
\newcommand{\PRA}[3]{\jcite{Phys. Rev. A}{#1}{#2}{#3}}
\newcommand{\PRB}[3]{\jcite{Phys. Rev. B}{#1}{#2}{#3}}
\newcommand{\jcc}[3]{\jcite{J. Comput. Chem.}{#1}{#2}{#3}}
\newcommand{\molphys}[3]{\jcite{Mol. Phys.}{#1}{#2}{#3}}
\newcommand{\mph}[3]{\jcite{Mol. Phys.}{#1}{#2}{#3}}
\newcommand{\cpc}[3]{\jcite{Comput. Phys. Commun.}{#1}{#2}{#3}}
\newcommand{\jcsfii}[3]{\jcite{J. Chem. Soc. Faraday Trans. II}{#1}{#2}{#3}}
\newcommand{\prsa}[3]{\jcite{Proc. Royal Soc. A}{#1}{#2}{#3}}
\newcommand{\jacs}[3]{\jcite{J. Am. Chem. Soc.}{#1}{#2}{#3}}
\newcommand{\ijqcs}[3]{\jcite{Int. J. Quantum Chem. Symp.}{#1}{#2}{#3}}
\newcommand{\ijqc}[3]{\jcite{Int. J. Quantum Chem.}{#1}{#2}{#3}}
\newcommand{\spa}[3]{\jcite{Spectrochim. Acta A}{#1}{#2}{#3}}
\newcommand{\tca}[3]{\jcite{Theor. Chem. Acc.}{#1}{#2}{#3}}
\newcommand{\tcaold}[3]{\jcite{Theor. Chim. Acta}{#1}{#2}{#3}}
\newcommand{\jpcrd}[3]{\jcite{J. Phys. Chem. Ref. Data}{#1}{#2}{#3}}

\draft
\title{
On the integration accuracy in molecular density functional theory 
calculations using Gaussian basis sets}
\author{Jan M.L. Martin*}
\address{Department of Organic Chemistry,
B
Kimmelman Building, Room 262,
Weizmann Institute of Science,
IL-76100 Re\d{h}ovot, Israel. {\rm E-mail:} {\tt comartin@wicc.weizmann.ac.il}
}
\author{Charles W. Bauschlicher, Jr.*}
\address{Space Technology Division,
Mail Stop 230-3,
NASA Ames Research Center,
Moffett Field, CA 94035-1000, USA}

\author{Alessandra Ricca}
\address{ELORET,
Mail Stop 230-3,
NASA Ames Research Center,
Moffett Field, CA 94035-1000, USA}
\date{{\em Comput. Phys. Commun.} MS PTA019; Received February 14, 2000; Accepted May 5, 2000}
\maketitle
\begin{abstract}
The sensitivity of computed DFT (Density Functional Theory) molecular 
properties (including energetics, geometries, vibrational frequencies,
and infrared intensities) 
to the radial and angular numerical integration grid
meshes, as well as to the partitioning scheme, is discussed for a
number of molecules using the Gaussian 98 program system.
Problems with typical production
grid sizes are particularly acute for third-row transition metal systems,
but may still result in qualitatively incorrect results for a molecule
as simple as CCH.  Practical recommendations are made
with respect to grid choices for the energy(+gradient) steps, as 
well as for the solution of the CPKS (Coupled Perturbed Kohn-Sham)
equations.
\end{abstract}


\section{Introduction and theoretical background}

Density Functional Theory (DFT)\cite{Koh65,Par89} has in recent years
become both a powerful and a very commonly used tool for studying 
molecular electronic structure.  In addition to usually reliable energetics,
the availability of analytical first and second derivatives for the DFT approaches
allows for easy calculation of geometric structure and vibrational frequencies.
\par

The rate-determining step in molecular DFT calculations using Gaussian basis sets
is usually the evaluation of various integrals over the exchange-correlation
functional and its geometric derivatives. In practice the three-dimensional
numerical integration is carried out as a weighted sum of 
three-dimensional integrations in polar coordinates centered on atoms $A$:
\begin{equation}
\int{F({\bf r})d\tau} = \sum_{A}^{{\rm atoms}}\sum_{g}^{{\rm grid}}{w_{gA}p_A({\bf r_g})F({\bf r}_{gA})}
\end{equation}
in which the ${\bf r}_g$ and $w_g$ are numerical quadrature abscissae and weights, respectively,
and the partitioning function $p_A$ 
satisfies $\sum_{A} p_A({\bf r})=1$ for all values of ${\bf r}$. In other words,
$p_A$ acts as a `fader function' between
grids centered on different atoms.

Quite separate from issues involving the basis set and mathematical form of the
exchange-correlation functional, which determine the {\em accuracy} of the calculation,
the choice of atomic integration grid and partitioning can have a profound impact
on its numerical {\em precision} (together with trivial parameters like convergence
criteria). Especially since an increasing proportion of users of quantum chemical
software in general --- and DFT methods in particular --- are experimental
chemists without prior background in electronic structure theory, the tendency
exists to simply rely on the built-in defaults of various codes.

In the present paper, we shall report a number of problems which
may arise in this fashion. Because of its great popularity and widespread
use, our discussion will focus on the Gaussian~98\cite{g98} program 
system. We wish to emphasize, however, that the issues raised in the present
work could arise with other DFT codes. We note that some codes try
to minimize possible numerical problems by the use of large default integration
grids and/or by variation of the grid mesh with the atom type. While the
use of very large grids certainly avoids numerical problems, it can lead
to prohibitively expensive calculations for larger systems. Therefore, 
it is of interest to find the smallest grid that will yield precise results
for molecular properties.

Before proceeding to discuss our results, we shall first briefly review
some relevant mathematical aspects.

\section{Mathematical background}

The atomic integration grid is built up naturally as the direct product of
radial and angular grids:
\begin{equation}
\sum_g{w_gF({\rm r}_g)}=\sum_a\sum_b{w_a w_b F(r_a,\theta_b,\phi_b)}
\end{equation}
The angular grid itself can be either a direct product of appropriate
one-dimensional Gauss quadrature grids (e.g., Gauss-Legendre or Gauss-Chebyshev),
or (more efficiently) can be a two-dimensional Lebedev grid\cite{Lebedev} which 
exactly integrates spherical harmonics up to a certain order. (In order to
integrate spherical harmonics up to order $L$ exactly, the  direct product of grids
with $(L+1)/2$ abscissae in $\theta$ and $L+1$ abscissae in $\phi$ requires $(L+1)^2/2$ angular
quadrature points,
compared to approximately $(L+1)^2/3$ for the corresponding Lebedev grid\cite{AdamPhD}.)

The radial grids discussed in the present paper are all of the 
Murray-Handy-Laming (MHL) variety\cite{Han93} (following ideas first proposed by
Boys and Handy\cite{BoyHan}), in which the infinite interval 
$r=[0,\infty]$ is remapped onto the finite interval $q=[0,1]$ 
by means of the change of variable $r=\alpha (q/(1-q))^m$; the
integration in $q$ is then approximated by Euler-Maclaurin
summation. (The latter is facilitated by the fact that $\left(\frac{d^n r}{d q^n}\right)$
vanishes at $q=1$ for all $n$, and at $q=0$ up to $n=3m-1$ inclusive.)
Empirically, $m=$2 is found to yield the best results\cite{Han93}, and
$\alpha$ should be dependent on the atomic radius for best results.
The following working equations arise for abscissae and weights:
\begin{eqnarray}
r_i &=& R (i/(N+1-i))^2\\
w_i &=& 2 R^3 (N+1) i^5/(N+1-i)^7
\end{eqnarray}
in which $N$ is the number of radial points and $R$ is taken 
empirically\cite{Han93} as one-half the Bragg radius of the
atom on which the grid is centered (except for hydrogen, where
$R$ is set equal to the Bohr radius).

For molecular calculations, the partitioning function $p({\bf r})$ introduced in
Eq.(1) needs to be defined. 
One common choice is due
to Becke\cite{Bec88b}. Consider the confocal elliptical coordinate
$\mu_{AB}=r_A-r_B/R_{AB}$ (in which $r_A$ and $r_B$ are distances
towards nuclei $A$ and $B$, and $R_{AB}$ represents the internuclear
distance), and define $p(\mu)=\frac{1}{2}[1-g(\mu)]$, where
$g(\mu)$ is the `cell function'. The extreme choice $g(\mu_{AB})$=$-1$ if $\mu_{AB}\leq 0$;
=+1 if $\mu_{AB}>0$ (i.e. a Heaviside-type step function) would actually represent
a scheme in which space is simply partitioned into Voronoi polyhedra: this will
obviously be a numerically unstable scheme, since molecules by definition have 
appreciable electron density $\rho$ (of which the integrand is a functional) 
near bond midpoints $\mu_{AB}=0$.
Becke\cite{Bec88b} proposed to use a smoother approximation 
$g(\mu)=f(f(f(\mu)))$, where
$f(\mu)=\frac{3}{2}\mu-\frac{1}{2}\mu^3$.
Stratmann, Scuseria, and Frisch (SSF)\cite{Scu96} proposed to use a
piecewise function instead, namely $g(\mu_{AB})$=$-1$ if $\mu_{AB}\leq -a$;
=+1 if $\mu_{AB}\geq a$;
$=\frac{1}{16}[35(\mu_{AB}/a)-35(\mu_{AB}/a)^3+21(\mu_{AB}/a)^5-5(\mu_{AB}/a)^7]$
otherwise, where the scale factor $a=0.64$ was determined empirically.\footnote{As pointed out by a referee, this latter expression is equivalent to 
eq. (24), with $m$=3, of Ref.\cite{Han93}, although a piecewise $g(\mu)$ was not proposed there.}
The SSF partitioning is reported to be numerically somewhat more stable, and
in addition lends itself well to linear-scaling implementations\cite{Scu96}.

\section{Computational methods}

All calculations were carried out using Gaussian~98 Rev. A7\cite{g98} running
on various workstation computers at NASA Ames Research Center and the
Weizmann Institute of Science.

A variety of Gaussian basis sets was used, including the Pople group 
6-31G*, 6-31+G*, and 6-31++G** basis sets\cite{Heh85}, the Dunning cc-pVDZ and cc-pVTZ 
basis sets (correlation consistent\cite{Dun89,ccecc} polarized valence double and triple
zeta, respectively), and the Hay-Wadt LANL2DZ (Los Alamos National Laboratory
2 Double-Zeta\cite{lanl}) basis set-relativistic ECP combination.

In virtually all cases, the very popular B3LYP (Becke 3-parameter-Lee-Yang-Parr)
hybrid exchange-correlation functional\cite{Bec93,Lee88} was used,
except for some test calculations using the earlier BLYP
(Becke-Lee-Yang-Parr\cite{Lee88,Bec88}) or
BP86 (Becke-Perdew 1986\cite{Bec88,Per86}) functionals.

We consider three types of grids:
\begin{itemize}
\item ($nn$,$mmm$), meaning
the direct product of an $nn$-point  Euler-Maclaurin radial grid with a
two-dimensional Lebedev grid with $mmm$ angular points;
\item ($nn$,$mm$,2$mm$), meaning the direct product of an Euler-Maclaurin 
grid with $nn$ points in $r$ and of Gauss-Legendre quadratures with $mm$ 
and $2mm$ points  in $\theta$ and $\phi$, respectively;
\item ($nn$,$mmm$)p, i.e. pruned ($nn$,$mmm$) grids. In pruned grids\cite{sg1},
the
number of angular points is varied with the radial coordinate, with the
full number of $mmm$ only being used at $r$ values relevant for 
conventional chemical bonding. In Gaussian~98, the following pruned 
grids are implemented: (35,110)p (invoked with the keyword ``Grid=Coarse''),
(50,194)p (the SG-1 grid\cite{sg1} invoked with the keyword ``Grid=SG1''),
(75,302)p (invoked with the keyword ``Grid=Fine''), and 
(99,590)p (invoked as ``Grid=UltraFine''). Pruned grids are only implemented
for elements up to and including krypton: for heavier elements, the
corresponding unpruned grid is substituted. (We note that in the implementation
in Gaussian~98, the cutoff radii between segments with 
different angular grid meshes are dependent on the row in the Periodic Table:
this in effect does result in some grid quality adjustment with the atomic 
number.)
\end{itemize}

In frequency calculations using analytical second derivatives\cite{Joh94}, 
Gaussian~98 permits the use of different grids for the evaluation 
of the energy and gradient on the one hand, and for the solution of
the CPKS (coupled perturbed Kohn-Sham) equations on the other hand.
The defaults of the program correspond to ``Int(Grid=Fine) 
CPHF(Grid=Coarse)''; specifying ``Int(Grid=UltraFine)'' implies
``CPHF(Grid=SG1)'' by default.\footnote{In Gaussian 98, the keyword CPHF
controls options for both the Coupled Perturbed Hartree-Fock procedure in
wavefunction based ab initio calculations and for CPKS in density 
functional calculations.}

Unless noted otherwise, tightened convergence criteria were used
in geometry optimizations.

\section{Results and discussion}

\subsection{CCH radical}

The acetyl radical is linear with a $^{2}\Sigma^{+}$ ground state.
At the B3LYP/cc-pVTZ level with the default grid combination, however,
one surprisingly finds (Table \ref{cch_tab}) 
not only an imaginary bending frequency, but that the two
``degenerate" components differ by 64~cm$^{-1}$.
Upon repeating the calculation with an extremely large (140,48,96) 
grid, a properly degenerate bending frequency of 312.5 
cm$^{-1}$ (intensity 3.84 km/mol) was found, irrespective of whether
Becke or SSF partitioning is used. Taking the SSF (140,48,96) result as the 
reference, we find the default grid to be in error by only 0.3 and 0.1
cm$^{-1}$ for the CC and CH stretches, respectively, while the two 
components of the bending frequency are off by a whopping $-421$ and 
$-492$~cm$^{-1}$, respectively. (N.b.: we have treated the imaginary
frequencies as negative numbers when we compute the error.)
Likewise, while errors of about 5 and 1 \%
are seen for the CC and CH intensities, respectively, the bend 
intensity is too high by an order of magnitude. Using Becke instead of 
SSF weights does not materially affect these observations.

Specifying ``Grid=Ultrafine'' (i.e. (99,590)p for integration, and
(50,194)p for CPKS) leads to a dramatic improvement in
the bending frequency, which is now properly degenerate and only 7 
cm$^{-1}$ too low. (The intensities for the other two modes now
agree with the reference values to two decimal places.)
The error in the corresponding infrared intensity 
drops by two orders of magnitude. Specifying the (99,590)p grid
for both integration and CPKS does not affect the stretching 
frequencies, but reduces the error on the bending frequency and 
intensity to 0.1 cm$^{-1}$ and 0.01 km/mol, respectively, using
either Becke or SSF partitioning. This clearly suggests that 
sensitivity is greatest to the CPKS grid.

Using the (75,302)p grid for integration and varying the CPKS grid
from the associated (35,110)p default, we find that the spurious
imaginary, nondegenerate bending frequencies disappear upon
using either the unpruned (35,110) or the larger pruned (50,194)p grid. 
However, errors of about --10 and --7~cm$^{-1}$ in the bending frequency remain, 
and are not reduced to the 1 cm$^{-1}$ range until a (75,302)p or even (75,302) CPKS grid is 
substituted. Upgrading the energy grid to (99,590)p does not
affect these conclusions. 

Surprisingly, using a (99,974) grid for both integration and CPKS
yields slightly different frequencies for the two components of
the $\pi$ bending mode.

  Upon finding an imaginary frequency, one is tempted to displace the
molecular geometry in the direction of that mode and reoptimize the
geometry.  Following this procedure, using the Gaussian~98 defaults (an energy grid of
(75,302)p, a CPKS grid of (35,110)p, and the SSF weights), one obtains
a HCC angle of 171.1~degrees and three real frequencies. However,
one of the six trivial vibrations (i.e., overall translation and rotation,
which should have exactly vanishing frequencies in an infinitely precise
calculation using a rotationally invariant method) has a frequency of
$508i$~cm$^{-1}$. Thus lowering symmetry has just made the numerical
problem less obvious.\par

  Unlike the frequencies which show a large sensitivity to the
choice of grid, the total energy of CCH is in error by less than 
1 microhartree for all grids studied.  Considering the small
effect on the energy, it is not surprising that the CCH results 
are most sensitive to the CPKS grid.

\subsection{Harmonic frequencies of TaCl$_{2}$}

The ground state for this symmetric linear molecule is 
$X~^{4}\Sigma^{-}_{g}$. Results at the B3LYP/LANL2DZ level
as a function of the integration and CPKS grids are summarized
in Table \ref{tacl2_tab}. Our largest grid (140,48,96) SSF results
will again be considered as the reference.

None of the grid combinations considered in Table \ref{tacl2_tab} lead to 
nonequivalent bending vibration components, but the default
grid combination does produce a spurious imaginary bending frequency.
Unlike CCH, the bending infrared intensities agree well with
the largest grid value, but the symmetric and antisymmetric stretching 
frequencies are in error by $-6$ and $-7$ cm$^{-1}$, respectively. 
Use of the Becke partitioning leads not only to increased errors in the 
stretching frequencies, but to an error of +0.002 \AA\ in the optimum
bond distance, compared to $-0.0003$~\AA\ with the SSF partitioning. 
Quite different from the behavior for CCH is the fact that the total
energy is now in error by 310 microhartree with the SSF partitioning,
and 369 microhartree with the Becke partitioning.

In this particular case, varying the size of the CPKS grid, and 
particularly of its angular component, does not appear to remedy
the problem at hand. Using an unpruned energy grid, however\footnote{
Note that this change will only affect the Cl atomic grid in this case, as grids
for elements heavier than Kr are always unpruned in Gaussian~98.},
or using a grid with more radial points, dramatically reduces the error
in the computed total energy, bond distance, and 
frequencies, particularly using the SSF partitioning. In particular,
the (99,590)p/(50,194)p integration/CPKS grid combination invoked
by ``Grid=UltraFine'' incurs (with SSF partitioning) 
no error greater than 1 cm$^{-1}$ on the frequencies, and reduces the
errors in bond distance and total energy to 0.0001 \AA\ and 1.5 
microhartree, respectively. While it is possible that using the Becke 
atomic size adjustment procedure (Appendix of Ref.\cite{Bec88b}), 
in conjuction 
with the SSF weights, would reduce the sensitivity 
to the radial grid, we note that using the
`UltraFine' grid combination 
solves both the CCH angular grid problem
and the TaCl$_2$ radial grid problem, and therefore would seem to be 
a good choice to invesigate possible grid problems.

For TaCl$_2$, performance of the SSF partitioning seems markedly superior
to that of the Becke partitioning. As expected, differences between 
the two partitioning schemes are reduced as the grid is improved, and
they yield essentially identical results for the (140,48,96) grid.

Lowering the symmetry to $C_{2v}$ and using the Gaussian~98 defaults
results in a slightly bent structure with a ClTaCl angle of 171.1~degrees.
As in the case of CCH, the only indication of the numerical problems
is in the deviation from zero of the translational and rotational 
frequencies, but for TaCl$_2$ the
value is $33i$~cm$^{-1}$, making the error for TaCl$_2$ less obvious
than in the case of CCH (see above).

If we now consider relative CPU times (default grids=1.00) as a 
function of grid size, we see that the  (99,590)p/(50,194)p 
combination will approximately double CPU time. Using a (99,590)p
grid for both integration and CPKS is found to be a factor of nine
more expensive than the default, a ratio which goes up to a factor
of 19 for the unpruned (99,974) grid, of 36 for the (96,32,64) grid
often cited in benchmarks, and of 109 for the largest (140,48,96)
grid considered here. The (99,590)p/(75,302)p combination found to
be required for 1 cm$^{-1}$ precision in the bending frequency of CCH
would be four times more expensive than the default --- rather more
time-consuming than desirable, but still an order of magnitude
less expensive than the (96,32,64) grid.

\subsection{Harmonic frequencies of Ge$_{2}$H$_{5}$ and 
Ge$_{2}$H$_{6}$}

   Ge$_2$H$_6$ has $D_{3d}$ symmetry and a structure analogous to ethane.  
Ge$_2$H$_5$ is best viewed as removing one H from Ge$_2$H$_6$, which
results in $C_s$ symmetry; there is only a small change in the structural 
parameters relative to the parent Ge$_2$H$_6$ molecule.  \par
\par
  In Table~\ref{t2} we summarize our results for Ge$_2$H$_5$. In addition
to the effect of partitioning and grid size, we also consider the influence of
the basis set and of the exchange-correlation functional on the three lowest harmonic frequencies of Ge$_2$H$_5$.
We first consider using the Becke partitioning.  Using the default grid in Gaussian,  i.e.
a pruned (75,302)p grid for energy and gradients and (35,110)p for CPKS,
results in an imaginary frequency. Expanding the radial grid from 75 to 96 points
and depruning the grid, i.e. the (96,302) grid, for the energy, gradient, and
CPKS steps does not remove the imaginary frequency.
Expanding the angular grid removes the numerical problem; with 590
angular points, the lowest harmonic frequency is essentially converged to
a  value of about 100~cm$^{-1}$.  Expanding the radial grid further has only
an effect of a few cm$^{-1}$.  
\par
   The results using SSF partitioning are very different from 
those obtained using Becke partitioning; using SSF, even the default ``Grid=Fine" is
precise to better than 10~cm$^{-1}$.  Improving
the grid makes much smaller changes for SSF and the 
``UltraFine" grid in Gaussian~98 (i.e., (99,590)p for energy and gradients, 
(50,194)p for CPKS) appears to be essentially converged. Note that
for the Becke partitioning the ``UltraFine" grid also yields essentially
converged  results.
\par
  Adding diffuse functions to the Ge atoms or diffuse and polarization functions
to the H atoms does not eliminate the imaginary frequency for the default grid
and the Becke partitioning.  That is, the problem is not unique to the
6-31G* basis set.\par
Nor is it unique to the use of a hybrid functional:
  substituting the BP86 functional for its B3LYP counterpart does not alter the
trends with respect to partitioning and grid,  the most notable point
being the imaginary frequency for the default grid with Becke partitioning.
\par
   The Ge$_2$H$_5$ calculations are repeated using the LANL2DZ basis set/ECP
combination,
and the results are also given in Table~\ref{t2}. Unlike the 
all-electron calculations, there is essentially no difference between
Becke and SSF partitioning, and improving the grid has only a very
small effect.  This is also true for the four- and 14-valence electron 
Ge ECPs of Hurley {\it et al.}\cite{ecp2}.  Thus it appears that the difference
observed in the all-electron calculations arises from the inner-shell orbitals.\par
  For the default grids and the Becke partitioning, displacing the molecular geometry
in the direction of the imaginary mode results in essentially the same
geometry and frequencies as found in the $C_s$ treatment.  That is, lowering the symmetry does not
eliminate the imaginary frequency, so the numerical problems
are much more obvious than for CCH and TaCl$_2$.\par
   Substituting Si for Ge leads to a similar structure, but as shown
in Table~\ref{t1}, neither the choice of grid points nor the
partitioning scheme make much of a difference for the Si$_2$H$_5$ system.  
The lack of sensitivity of the results to grid or partitioning is 
similar to the results reported\cite{Scu96}  by SSF, and is typical of many other 
tests that we have performed for systems containing first and second row atoms.  
The problem of the imaginary frequency does not exist for Ga$_2$Cl$_5$: the Becke and SSF partitioning
agree to within 4~cm$^{-1}$ for the 6-31+G* basis set and the default grids.
Given that Si$_2$H$_5$, Ge$_2$Cl$_5$, and Ge$_2$H$_5$  have similar shapes,
we conclude that the shape of Ge$_2$H$_5$ does not lead to the numerical problems
that result in the imaginary frequency.  \par

   In Table~\ref{t3} we summarize the results for Ge$_2$H$_6$, which
is a closed-shell, stable molecule.  It shows the same variation
in the frequencies with choice of partitioning and grid points as Ge$_2$H$_5$, and 
we conclude that the difference between results using Becke and SSF partitioning
is not unique to Ge$_2$H$_5$.  It is encouraging that the default SSF
partitioning is less sensitive to the choice of grid, but it is also
sobering to realize that it is possible to have an error of about 
70~cm$^{-1}$ with no obvious indication of an error.
\par

\subsection{Relative energies of cis- and trans-HIr[PR$_{3}$]$_{4}$Cl$^{+}$}

The peculiar cis-trans equilibrium of the 
HIr[PR$_{3}$]$_{4}$Cl$^{+}$ (R=H, CH$_{3}$)
transition metal complexes was the subject of a very recent joint
theoretical-experimental investigation\cite{isom}. At the highest level
of theory for R=CH$_3$, the complexes are found to be essentially 
isoenergetic, with the equilibrium between the two being essentially
wholly driven by the higher entropy of the cis form: 
experimentally\cite{isom}, a 6:1 cis:trans equilibrium is found at room
temperature.

Given the very small energy differences being considered here, as well 
as the rather flat potential surface of particularly the cis form,
grid  convergence (not discussed in Ref.\cite{isom}) 
was obviously a cause for concern here. 
Upon first optimizing the cis-R=CH$_3$ case using the default grids and 
computing vibrational frequencies, we found what appeared to be 
a local minimum. Upon recomputing the harmonic frequencies with the
(99,590)p/(50,194)p integral/CPKS grid combination, we found this 
structure to be a saddle point of order 3, and only after thirteen 
geometry optimization cycles using analytical second derivatives (and two
months of CPU time on an SGI Origin 2000) was the true minimum 
structure found. A superimposition of the two 
structures (created using the MOLDEN\cite{molden} molecular viewing
program) is displayed in Figure 1.

While the R=CH$_3$ case is simply too large to do an exhaustive grid 
convergence study, we have summarized some data for the R=H case
in Table \ref{ir_tab}. In particular, we have considered the absolute
energies for the cis and trans isomers, and their relative energy, as
a function of grid size. In this case, results seems to be quite 
dependent both on the radial and on the angular grid mesh. Using the
(35,110)p grid employed by default in the CPKS step, the relative 
energy gets predicted with the wrong sign, and neither the use of an
unpruned grid nor switching to a product grid of the same order
remedy the problem --- if anything, results get {\em worse}, suggesting
that the (35,110)p results in fact benefit from a fortunate error 
compensation. Total energies are off by as much as 8 {\em 
milli}hartree for the trans isomer. 
While a scatter in the energies does not necessarily imply a corresponding
scatter in the CPKS solution, it does not seem entiely implausible that a grid that 
predicts isomer energy differences on a surface with the wrong sign might
yield a Hessian with incorrect curvature if used in the CPKS step.

The smallest grid that at least predicts the correct sign for the
cis-trans difference is (50,194)p (about 25\% too high in absolute
value): using an unpruned grid instead\footnote{
Again, since the Ir grid is always unpruned, the improvement in grid quality
comes from additional grid points centered on the other atoms.
} cuts the error in half, as does
using the (75,302)p grid. Total energies are however still in error
by amounts of about 60 microhartree: fortunately the errors largely 
cancel. The (99,590)p (``UltraFine'') grid agrees to two decimal
places with the (140,48,96) reference results; even a grid as large
as (99,974) cannot yet achieve 1 microhartree precision in the total
energies, which is barely reached for a (99,27,54) grid, and more comfortably
for a (99,32,64) grid.

\section{Conclusions}

Computed molecular properties using density functional theory in 
general, and vibrational frequencies in particular, tend to be 
quite sensitive to the integration grids and partitioning
being used in the calculation.  Our results corroborate the assertion 
of SSF that their proposed partitioning scheme is numerically more 
stable than the Becke partitioning.

While the use of coarse grids in the solution of the CPKS (coupled perturbed
Kohn-Sham) equations certainly leads to significant speedups, it can
cause spurious negative eigenvalues in the Hessian matrix, or even 
artifactual loss of vibrational mode degeneracy, in systems as small 
and uncomplicated as CCH radical. While this problem is alleviated by
the use of unpruned grids, this offers no benefit (in terms of CPU 
time) over the use of a larger pruned grid. For systems involving 
very heavy elements (e.g. third-row transition metals), grids
which are sufficiently fine-meshed for typical first-and 
second-row systems are no longer adequate. In situations 
where small pruned grids are inadequate, the use of the corresponding
unpruned grid appears to be less efficient than the use
of a larger pruned grid.

For the cases considered here, the most efficient grid combination
that appears to be `immune' to qualitatively incorrect results is
(99,590)p for energy and gradients, (50,194)p for CPKS. (This is
equivalent to the ``Grid=UltraFine'' option in Gaussian~98.)
For greater quantitative precision, the use of a finer-meshed
(75,302)p grid for the CPKS is recommended. While a
(99,974), (96,32,64) or even finer grid in all steps of the 
calculation will certainly avoid this type of problems, this
comes at an unacceptably high premium in terms of CPU time.

As quantum chemical methods mature and quantum chemical software
becomes more user-friendly, an increasing proportion of their
users are scientists from other fields than quantum chemistry, 
who are inevitably exposed to the temptation to treat such
program systems as `black boxes'. Observations such as those
made in the present paper illustrate that even at the present
state of technology, `black box' performance should not be
taken for granted, and that acquiring some insight in both
the quantum mechanical and numerical methodology is still
an essential prerequisite for  their effective and reliable 
scientific application.

\acknowledgments

JM is the incumbent of the Helen and Milton A. Kimmelman Career 
Development Chair.
Research at the Weizmann Institute 
was supported by the Minerva Foundation, Munich, Germany, and
by the {\em Tashtiyot} program of the Ministry of Science (Israel).
JM would like to thank Dr. S. Parthiban for bringing the problem 
with CCH to his attention,
and Dr. Andreas Sundermann for critical reading of the draft manuscript.
AR was supported by NASA Contract No. NAS2-14031.\par

\section*{Note added}

After acceptance of the present manuscript, and in response to a preprint
thereof, we received a personal
communication from Gary W. Trucks and Michael J. Frisch of Gaussian,
Inc., informing us of a bug in Gaussian 98 which affects the integral
(derivative) accuracy for the {\em specific} case of the (35,110)p grid.
A fix will be incorporated in the next minor release after Rev.A9.
The main effect on the data presented in this paper is that the
errors in the first two entries in Table 1 become a good deal less
dramatic: in particular, the degeneracy of the bend is restored and
the errors in $\omega_3$ and $I_3$ are reduced to -19 cm$^{-1}$ and
1.2 km/mol, respectively.  In addition, we reoptimized the complex in 
Figure 1 using a patched version of the code with the default grid
combination (and using analytical second derivatives as before). The
resulting geometry is in much closer agreement with the Grid=UltraFine
structure. (The cis-trans difference with the (35,110)p grid in 
Table VI is not affected materially.)
The authors would like to 
thank Drs.  Trucks and Frisch for helpful discussions.

\begin{table}
\caption{\label{cch_tab}Numerical errors in computed B3LYP/cc-pVTZ properties for acetyl
radical as a function of grid size.  The values are given for the largest grid, which is used
as our reference.}
\squeezetable
\begin{tabular}{lllddddddddd}
Energy & CPKS  &  Partition  &  $\Delta$E(uEh)  &  
$\Delta\omega_{3x}$  &  $\Delta I_{3x}$  &  $\Delta\omega_{3y}$  &  $\Delta I_{3y}$  &  $\Delta\omega_{2}$  &  
$\Delta I_{2}$  & $\Delta\omega_{1}$  &  $\Delta I_{1}$\\
grid & grid  &  scheme  &  $\mu E_h$  &  cm$^{-1}$  &  km/mol  &  cm$^{-1}$  &  km/mol  &  cm$^{-1}$  &  km/mol  &  cm$^{-1}$  &  km/mol\\
\hline
(75,302)p  & (35,110)p   &  B  &  0.65  &  -481.35  &  38.53  &  -417.12  &  30.42  &  -0.21  &  0.28  &  0.62  &  -0.53\\
(75,302)p  & (35,110)p   &  SSF  &  -0.10  &  -492.09  &  39.55  &  -420.61  &  30.92  &  0.29  &  0.28  &  0.08  &  -0.55\\
(75,302)p  & (35,110)    &  B  &  0.65  &  -10.07  & 0.61 &-10.07  &
0.61 &
-0.32 & 0.00 & 0.44 & 0.00\\
(75,302)p  & (35,110)    & SSF &  -0.10 &   -9.42  & 0.61 & -9.42  &
0.61 &
0.17 & 0.00 & -0.12 & -0.01\\
(75,302)   & (35,110)    &  B  &  0.59 & -9.84 & 0.61 & -9.84 & 0.61 & -0.32 & -0.01 & 0.45 & 0.00\\
(75,302)   & (35,110)    &  SSF  &  -0.16  &  -9.25  &  0.61  &  -9.25  &  0.61  &  0.17  &  0.00  &  -0.11  &  0.00\\
(75,302)p  & (50,194)p   &  B  &  0.65  &  -7.43  &  0.44  &  -7.43  &  0.44  &  -0.32  &  -0.01  &  0.44  &  0.00\\
(75,302)p  & (50,194)p   &  SSF  &  -0.10  &  -6.79  &  0.44  &  -6.79  &  0.44  &  0.17  &  0.00  &  -0.12  &  -0.01\\
(75,302)p  & (75,302)p   &  B  &  0.65  &  0.73  &  -0.08  &  0.73  &  -0.08  &  -0.32  &  -0.01  &  0.44  &  0.00\\
(75,302)p  & (75,302)p   &  SSF  &  -0.10  &  1.34  &  -0.08  &  1.34  &  -0.08  &  0.17  &  0.00  &  -0.12  &  -0.01\\
(75,302)   & (75,302)    &  B  &  0.59  &  0.08  &  -0.02  &  0.08  &  -0.02  &  -0.32  &  0.00  &  0.45  &  0.00\\
(75,302)   & (75,302)    &  SSF  &  -0.16  &  0.64  &  -0.02  &  0.64  &  -0.02  &  0.17  &  0.00  &  -0.11  &  0.00\\
(99,590)p  & (50,194)p   &  B   &  -0.04  &  -7.06  &  0.45  &  -7.06  &  0.45  &  0.06  &  0.00  &  0.10  &  0.00\\
(99,590)p  & (50,194)p   &  SSF  &  0.12  &  -7.06  &  0.45  &  -7.06  &  0.45  &  -0.09  &  0.00  &  -0.29  &  0.00\\
(99,590)p  & (75,302)p   &  B  &  -0.04  &  1.08  &  -0.07  &  1.08  &  -0.07  &  0.06  &  0.00  &  0.10  &  0.00\\
(99,590)p  & (75,302)p   &  SSF  &  0.12  &  1.09  &  -0.07  &  1.09  &  -0.07  &  -0.09  &  0.00  &  -0.29  &  0.00\\
(99,590)p  & (99,590)p   &  B  &  -0.04  &  0.11  &  -0.01  &  0.11  &  -0.01  &  0.06  &  0.00  &  0.10  &  0.00\\
(99,590)p  & (99,590)p   &  SSF  &  0.12  &  0.12  &  -0.01  &  0.12  &  -0.01  &  -0.09  &  0.00  &  -0.29  &  0.00\\
(99,590)   & (99,590)    &  B  &  -0.04  &  0.07  &  -0.01  &  0.07  &  -0.01  &  0.06  &  0.00  &  0.10  &  0.00\\
(99,590)   & (99,590)    &  SSF  &  0.12  &  0.07  &  -0.01  &  0.07  &  -0.01  &  -0.09  &  0.00  &  -0.29  &  0.00\\
(99,974)   & (99,974)    &  B  &  -0.06  &  0.01  &  0.00  &  0.73  &  0.00  &  0.06  &  0.00  &  0.08  &  0.00\\
(99,974)   & (99,974)    &  SSF  &  0.11  &  0.03  &  0.00  &  0.73  &  0.00  &  -0.08  &  0.00  &  -0.25  &  0.00\\
(99,36,72) &(99,36,72) &  B   &  -0.07  &  0.14  &  -0.01  &  0.14  &  -0.01  &  0.06  &  0.00  &  0.08  &  0.00\\
(99,36,72) &(99,36,72) &  SSF  &  0.11  &  0.15  &  -0.01  &  0.15  &  -0.01  &  -0.08  &  0.00  &  -0.25  &  0.00\\
(140,48,96)&(140,48,96)&  B   &  0.00  &  0.01  &  0.00  &  0.01  &  0.00  &  0.02  &  0.00  &  -0.01  &  0.00\\
(140,48,96) & (140,48,96) & SSF & ...$^a$ & 312.54 & 3.85 & 312.54 & 3.85 & 2094.63 & 5.05 & 3456.85 & 56.60\\
\end{tabular}
\noindent
$^a$  The total energy is -76.63761769~Hartrees.\par
\end{table}

\begin{table}
\caption{\label{tacl2_tab} Numerical errors in computed B3LYP/LANL2DZ properties of 
TaCl$_2$ ($X~^{4}\Sigma^{-}_{g}$ state) as a function of grid size.
The values are given for the largest grid, which is used as our reference.}
\squeezetable
\begin{tabular}{llldddddddd}
Energy  & CPKS   &Partition&  $\Delta$r  &  $\Delta$E  &  $\Delta\omega_3$  &  
$\Delta I_3$  &   $\Delta\omega_2$  & $\Delta\omega_1$  &   $\Delta I_1$ & relative\\
Grid   & Grid  &scheme &  m\AA  &  $\mu E_h$  &  cm$^{-1}$  &  km/mol  &    cm$^{-1}$  &  cm$^{-1}$  &  km/mol  &  CPU time\\
  \hline
(75,302)p  &  (35,110)p  &  B  &  1.86  &  368.55  &  -60.21  &  0.02  &  -7.67  &  -8.86  &  -0.23  &  1.00\\
(75,302)p  &  (35,110)p  &  SSF  &  -0.26  &  309.60  &  -59.90  &  -0.01  &  -5.40  &  -6.80  &  -0.01  &  1.00\\
(75,302)  &  (35,110)  &  B  &  0.04  &  -17.03  &  3.97  &  0.00  &  0.15  &  0.20  &  -0.01  &  3.05\\
(75,302)  &  (35,110)  &  SSF  &  0.11  &  -5.48  &  1.83  &  0.00  &  -0.15  &  -0.16  &  -0.01  &  2.98\\
(75,302)p  &  (50,194)p  &  B  &  1.86  &  368.55  &  -60.25  &  0.02  &  -7.67  &  -8.85  &  -0.22  &  2.02\\
(75,302)p  &  (50,194)p  &  SSF  &  -0.26  &  309.60  &  -59.95  &  -0.01  &  -5.40  &  -6.80  &  0.00  &  2.01\\
(75,302)p  &  (75,302)p  &  B  &  1.86  &  368.55  &  -60.26  &  0.02  &  -7.67  &  -8.86  &  -0.22  &  3.86\\
(75,302)p  &  (75,302)p  &  SSF  &  -0.26  &  309.60  &  -59.95  &  -0.01  &  -5.41  &  -6.80  &  0.00  &  3.82\\
(99,590)p  &  (35,110)p  &  B  &  0.09  &  -16.84  &  4.52  &  0.00  &  0.15  &  0.23  &  -0.02  &  1.16\\
(99,590)p  &  (35,110)p  &  SSF  &  0.10  &  -1.54  &  0.91  &  0.00  &  -0.06  &  -0.04  &  -0.02  &  1.15\\
(99,590)p  &  (50,194)p  &  B  &  0.09  &  -16.84  &  4.49  &  0.00  &  0.15  &  0.23  &  -0.01  &  2.19\\
(99,590)p  &  (50,194)p  &  SSF  &  0.10  &  -1.54  &  0.88  &  0.00  &  -0.06  &  -0.04  &  -0.01  &  2.17\\
(99,590)p  &  (75,302)p  &  B  &  0.09  &  -16.84  &  4.49  &  0.00  &  0.15  &  0.23  &  -0.01  &  4.00\\
(99,590)p  &  (75,302)p  &  SSF  &  0.10  &  -1.54  &  0.88  &  0.00  &  -0.06  &  -0.04  &  -0.01  &  3.97\\
(99,590)p  &  (99,590)p  &  B  &  0.09  &  -16.84  &  4.49  &  0.00  &  0.15  &  0.23  &  -0.01  &  8.96\\
(99,590)p  &  (99,590)p  &  SSF  &  0.10  &  -1.54  &  0.88  &  0.00  &  -0.06  &  -0.04  &  -0.01  &  8.91\\
(99,974)  &  (99,974)  &  B  &  0.00  &  0.11  &  -0.02  &  0.00  &  -0.01  &  -0.01  &  0.00  &  19.2\\
(99,974)  &  (99,974)  &  SSF  &  0.00  &  0.14  &  0.00  &  0.00  &  -0.01  &  -0.02  &  0.00  &  19.1\\
(96,32,64)  &  (96,32,64)  &  B  &  0.00  &  -0.01  &  0.11  &  0.00  &  -0.02  &  -0.03  &  0.00  &  36.7\\
(96,32,64)  &  (96,32,64)  &  SSF  &  0.00  &  0.28  &  0.00  &  0.00  &  -0.05  &  -0.06  &  0.00  &  36.3\\
(99,48,96)  &  (99,48,96)  &  B  &  0.00  &  0.08  &  0.00  &  0.00  &  -0.01  &  -0.01  &  0.00  &  80.1\\
(99,48,96)  &  (99,48,96)  &  SSF  &  0.00  &  0.14  &  0.00  &  0.00  &  -0.01  &  -0.02  &  0.00  &  79.1\\
(140,48,96)  &  (140,48,96)  &  B  &  0.00  &  0.00  &  0.01  &  0.00  &  0.00  &  0.00  &  0.00  &  110.4\\
(140,48,96) & (140,48,96) & SSF & 2335.357 & ...$^a$ & 35.89 & 3.96 &347.51 & 375.63 & 107.73& 109.4\\
\end{tabular}
\noindent
$^a$ The total energy is --87.74246174~Hartrees.\par

\end{table}

\begin{table}
\caption{\label{t2} Summary of the three lowest harmonic frequencies of
Ge$_2$H$_5$. }
\begin{center}
\begin{tabular}{llrrrrrr}
Energy & CPKS& \multispan 3 \hfil  Becke partitioning \hfil & \multispan 3 
\hfil SSF partitioning \hfil \\
Grid & Grid & \multispan 3 \hrulefill & \multispan 3 \hrulefill \\
\multispan 8 \hfil B3LYP/6-31G* \hfil \\
(75,302)p& (35,110)p &  77.9\rlap{i}& 250.2& 352.2& 114.2& 250.5& 363.6\\
(75,302)&  (75,302)  &  77.6\rlap{i}& 250.2& 352.2& 114.2& 250.5& 363.6\\
(96,302)&   (96,302) &  39.6\rlap{i}& 250.2& 356.0& 101.0& 250.4& 361.5\\
(75,434)&   (75,434) &  44.6        & 250.4& 365.5& 107.4& 250.4& 358.4\\
(75,590)&  (75,590)  & 108.4        & 250.3& 360.6&  95.3& 250.4& 360.4\\
(99,590)p& (50,194)p & 106.4        & 250.4& 359.9& 103.1& 250.4& 360.0\\
(99,590)&  (99,590)  & 106.5        & 250.4& 359.9& 103.1& 250.4& 359.9\\
(75,770)&  (75,770)  & 103.6        & 250.3& 360.5& 105.8& 250.4& 361.5\\
(75,974)&  (75,974)  & 107.7        & 250.3& 360.2& 103.9& 250.4& 360.5\\
(96,974)&  (96,974)  & 106.2        & 250.3& 360.5& 105.0& 250.4& 360.3\\
(128,974)& (128,974) & 106.8        & 250.4& 360.3& 105.2& 250.4& 360.1\\
\multispan 8 \hfil B3LYP/6-31+G* \hfil \\
(75,302)p& (35,110)p &  82.7\rlap{i}& 250.7& 352.8& 111.6& 250.8& 364.1\\
\multispan 8 \hfil B3LYP/6-31++G** \hfil \\
(75,302)p& (35,110)p &  36.6\rlap{i}& 252.8& 349.7& 110.9& 252.8& 368.8\\
\multispan 8 \hfil BP86/6-31G* \hfil \\
(75,302)p& (35,110)p & 109.8\rlap{i}& 247.4& 337.7&  95.4& 247.9& 349.1\\
(99,590)& (99,590)&    108.0        & 247.9& 347.5&  96.9& 247.9& 346.9\\
\multispan 8 \hfil B3LYP/LANL2DZ \hfil \\
(75,302)p& (35,110)p &  88.8        & 241.5& 373.7&  89.1& 241.5& 373.7\\
(99,590)&   (99,590) &  88.3        & 241.5& 373.8&  88.3& 241.5& 373.8\\
\end{tabular}
\end{center}
\noindent
\end{table}
\clearpage
\begin{table}
\caption{\label{t1} Summary of the three lowest and the three highest
harmonic frequencies of Si$_2$H$_5$ using the 6-31G* basis set and
the B3LYP functional. }
\begin{center}
\begin{tabular}{lllrrrrrr}
Energy& CPKS& Partition& \multispan 6 \hfil Harmonic frequencies \hfil \\
Grid  & Grid& Scheme& \multispan 6 \hrulefill \\
(75,302)p& (35,110)p & B   & 123.4& 390.4& 407.1& 2233.2& 2244.8& 2254.5\\
(75,302)p& (35,110)p & SSF & 124.0& 390.2& 407.2& 2227.9& 2241.1& 2250.8\\
(75,302)& (75,302)& B      & 123.3& 390.4& 407.0& 2233.2& 2244.8& 2254.5\\
(75,302)& (75,302)& SSF    & 123.9& 390.3& 407.1& 2227.9& 2241.1& 2250.8\\
(96,974)&   (96,974)&   B  & 122.2& 390.6& 408.6& 2231.2& 2243.5& 2253.2\\
(96,974)&  (96,974)&  SSF  & 122.1& 390.6& 408.8& 2229.9& 2242.8& 2252.4\\
(99,590)p& (50,194)p & B   & 125.4& 391.2& 409.8& 2231.1& 2243.3& 2253.0\\
(99,590)p& (50,194)p & SSF & 123.1& 390.7& 409.0& 2230.8& 2243.2& 2252.8\\
\end{tabular}
\end{center}
\end{table}
\clearpage
\begin{table}
\caption{\label{t3} Summary of the three lowest harmonic frequencies of Ge$_2$H$_6$
using the 6-31G* basis set and the B3LYP functional. }
\begin{center}
\begin{tabular}{llrrrrrrr}
Energy & CPKS& \multispan 3 \hfil  Becke partitioning \hfil & \multispan 3 
\hfil  SSF partitioning \hfil
 \\
Grid & Grid & \multispan 3 \hrulefill & \multispan 3 \hrulefill \\
(75,302)p& (35,110)p &  57.1        & 259.7& 347.0& 123.9& 259.8& 360.7\\
(75,302)&  (75,302)  &  54.1        & 259.7& 346.8& 122.6& 259.8& 360.5\\
(75,434)&  (75,434)  &  44.6\rlap{i}& 259.7& 351.7& 116.8& 259.8& 353.0\\
(75,590)&  (75,590)  &  93.0        & 259.7& 347.5& 111.7& 259.8& 356.3\\
(99,590)p& (50,194)p &  94.2        & 259.8& 347.3& 107.7& 259.8& 353.2\\
(99,590)&  (99,590)  &  94.5        & 259.8& 347.3& 107.9& 259.8& 353.2\\
(75,770)& (75,770)   & 106.6        & 259.7& 353.9& 112.3& 259.8& 355.1\\
(75,974)& (75,974)   & 110.7        & 259.7& 353.6& 108.6& 259.8& 354.1\\
(96,302)&  (96,302)  &  75.2        & 259.7& 351.9& 109.9& 259.8& 356.3\\
(96,434)&  (96,434)  &  30.0\rlap{i}& 259.8& 351.8& 102.6& 259.8& 352.9\\
(96,974)& (96,974)   & 110.1        & 259.7& 353.8& 109.0& 259.8& 353.8\\
(128,974)& (128,974) & 110.3        & 259.8& 353.8& 109.4& 259.8& 353.7\\
\end{tabular}
\end{center}
\end{table}


\begin{table}
\caption{\label{ir_tab}Numerical errors in absolute and relative
B3LYP/LANL2DZ
energies of cis and trans isomers$^a$ of HIr(PH$_3$)$_4$Cl$^+$ as
a function of grid size}
\squeezetable
\begin{tabular}{lddd}
Grid        &   $\Delta E$ &$\Delta E$ & Isomerization \\
            &   trans($C_{4v}$)      &  cis($C_s$)      & energy \\
            & $\mu E_h$  & $\mu E_h$  & kcal/mol\\
\hline
(35,110)p  &  2488.42  &  890.21  &  0.409\\
(35,110)  &  5638.22  &  1301.61  &  2.127\\
(35,9,18)  &  8400.38  &  113.28  &  4.606\\
(50,194)p  &  -236.21  &  112.62  &  -0.813\\
(50,194)  &  -118.27  &  17.55  &  -0.679\\
(50,12,24)  &  653.35  &  77.73  &  -0.233\\
(50,302)  &  -110.06  &  17.55  &  -0.674\\
(50,15,30)  &  -118.27  &  -95.50  &  -0.608\\
(75,302)p  &  -58.35  &  63.27  &  -0.670\\
(75,302)  &  -33.64  &  12.74  &  -0.623\\
(75, 15, 30)  &  -23.58  &  -44.17  &  -0.581\\
(99,194)  &  -68.07  &  52.53  &  -0.670\\
(99,12,24)  &  708.68  &  135.89  &  -0.235\\
(99, 302)  &  -33.52  &  10.52  &  -0.622\\
(99, 15, 30)  &  -21.44  &  -44.17  &  -0.580\\
(99, 590)p  &  -15.29  &  -7.81  &  -0.599\\
(99, 590)  &  -10.22  &  -5.75  &  -0.597\\
(99, 21, 42)  &  0.68  &  5.52  &  -0.597\\
(99, 974)  &  -2.29  &  0.70  &  -0.596\\
(99, 27, 54)  &  0.39  &  -0.97  &  -0.593\\
(99, 32, 64)  &  0.64  &  0.21  &  -0.594\\
(99, 36, 72)  &  0.11  &  0.13  &  -0.594\\
(50, 48, 96)  &  -85.20  &  -45.60  &  -0.619\\
(75, 48, 96)  &  -2.03  &  2.24  &  -0.597\\
(99, 48, 96)  &  -0.13  &  0.14  &  -0.594\\
(140, 48, 96)  &  0.00$^b$  &  0.00$^b$  &  -0.594\\
(50, 590)  &  -92.82  &  -49.43  &  -0.621\\
(75, 590)  &  -12.14  &  -3.19  &  -0.600\\
(99, 590)  &  -10.22  &  -5.75  &  -0.597\\
(120, 590)  &  -10.19  &  -5.89  &  -0.597\\
\end{tabular}
\noindent
$^a$ Reference geometries were obtained at the B3LYP/LANL2DZ level with
the (99,590)p/(50,194)p grid combination and tightened optimization
criteria.\par
$^b$ Total energies: cis $-$153.25208234, trans $-$153.25302915 hartree.\par

\end{table}

\clearpage
\begin{figure}
\caption{Superimposition of B3LYP/LANL2DZ 
optimized structures for cis-HIr[P(CH$_3$)$_3$]$_4$Cl$^+$
with the (75,302)p/(35,110)p (dark grey)
and (99,590)p/(50,194)p (light grey) grid combinations.}
\vspace*{48pt}
\includegraphics[bb=48 300 455 740, clip, angle=90, scale=1.0]{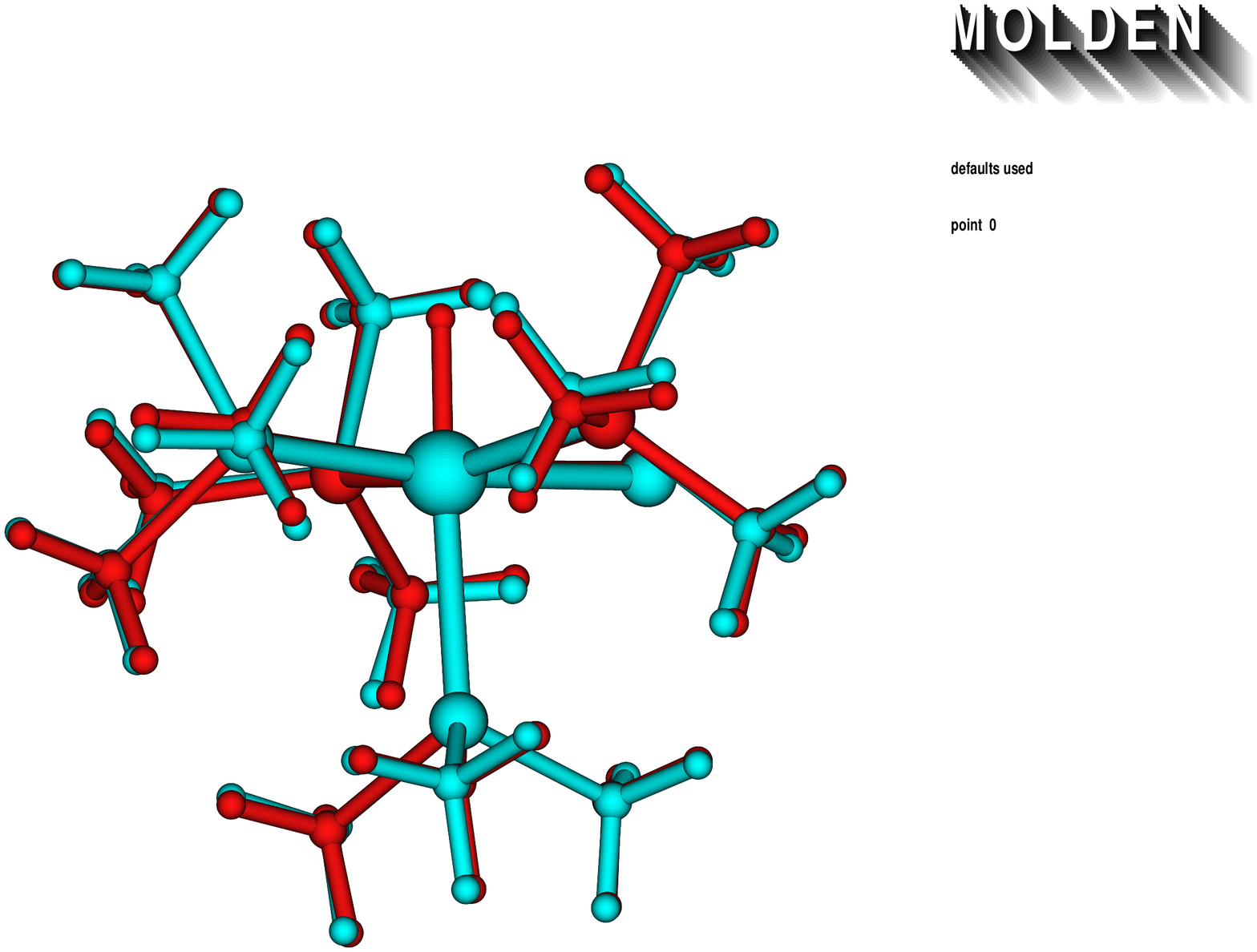}\\
\end{figure}
\end{document}